%% file: main.tex
\documentclass[sigconf,natbib=true,anonymous=false,balance]{acmart}
\usepackage[utf8]{inputenc}
\usepackage{booktabs}
\usepackage{multirow}
\usepackage{ctable}
\usepackage{tabularx}
\usepackage{graphicx}
\usepackage{array}
\usepackage[all]{nowidow}
\usepackage{balance}
\usepackage{xspace}
\usepackage{url}
\usepackage{fixltx2e}
\usepackage{caption,subcaption}
\usepackage{commath}
\usepackage{breqn}

\input{macros}

\input{orenLocalTemplates}

\copyrightyear{2022}
\acmYear{2022}
\setcopyright{acmlicensed}\acmConference[SIGIR '22]{Proceedings of the 45th International ACM SIGIR Conference on Research and Development in Information Retrieval}{July 11--15, 2022}{Madrid, Spain}
\acmBooktitle{Proceedings of the 45th International ACM SIGIR Conference on Research and Development in Information Retrieval (SIGIR '22), July 11--15, 2022, Madrid, Spain}
\acmPrice{15.00}
\acmDOI{10.1145/3477495.3531727}
\acmISBN{978-1-4503-8732-3/22/07}

\begin{document}
\fancyhead{}
\title{A Dataset for Sentence Retrieval for Open-Ended Dialogues }

\author{Itay Harel}
\authornote{Work done while at the Technion.}
\affiliation{%
  \institution{TSG IT Advanced Systems Ltd.}
  \city{Tel Aviv}
  \country{Israel}}
\email{itay.harel91@gmail.com}

\author{Hagai Taitelbaum}
\affiliation{%
  \institution{Google Research}
  \city{Tel Aviv}
  \country{Israel}}
\email{hagait@google.com}

\author{Idan Szpektor}
\affiliation{%
  \institution{Google Research}
  \city{Tel Aviv}
  \country{Israel}}
\email{szpektor@google.com}

\author{Oren Kurland}
\affiliation{%
  \institution{Technion {\textemdash} Israel institute of technology}
  \city{Haifa}
  \country{Israel}}
\email{kurland@ie.technion.ac.il	}

\input{category}
\input{abstract}
\maketitle
%\hagai{I think the template corresponds to a conference from July 2017}
%\oren{Added the latest template}
%\hagai{Thanks! Look at the top left corner in the next page, it's still 2017, I'm not sure how to change it and to what}
%\oren{It doesn't matter; this is their default}
\input{intro}

\input{rel}
\input{dataset}
\input{models}

\input{experiments}

\input{conc}

%\begin{figure}[t]
% \begin{figure*}[ht]
%     \makebox[1 \columnwidth][c]{
%     \includegraphics[width=1\columnwidth, height=7cm]{./images/guidelines_to_workers}
%     }
%     \caption{The test set annotation guidelines for MTurk workers.}
%     \label{fig:mturk_guidelines}
% %\end{figure}

% \end{figure*}

\input{Acknowledgements}

\appendix
\input{appendix}

\bibliographystyle{ACM-Reference-Format}
\bibliography{references,orenBib}

\end{document}

%% file: macros.tex
\newcommand{\omt}[1]{}

\newcommand{\myparagraph}[1]{\vspace{0.3\baselineskip}\noindent{\textbf{#1}}.~}
\newcommand{\doc}{d}

\newcommand{\arbTerm}{w}

\newcommand{\BERTfineMS}{MS\xspace}
\newcommand{\BERTfineR}{MS$\rightarrow$R\xspace}

\renewcommand{\set}[1]{\{#1\}}
\newcommand{\definedas}{\stackrel{def}{=}}

\newcommand{\crossEnt}{CE\xspace}
\newcommand{\indSet}{{\mathcal I}}
\newcommand{\okapi}{Okapi\xspace}
\newcommand{\ce}[2]{\crossEnt(#1 \; \vert\vert \,\, #2)}

\newcommand{\map}{MAP\xspace}

\newcommand{\mrr}{MRR\xspace}
\newcommand{\ndcg}{\ndcgfive\xspace}
\newcommand{\ndcgfive}{NDCG@5\xspace}

\newcommand{\asker}{initiator\xspace}
\newcommand{\answerer}{responder\xspace}
\newcommand{\answerers}{responders\xspace}

\newcommand{\initRank}{Initial Ranker\xspace}

%% file: orenLocalTemplates.tex
\newcommand{\post}{turn\xspace}
\newcommand{\posts}{turns\xspace}
\newcommand{\target}{target \post}
\newcommand{\targets}{target \posts}
\newcommand{\ut}{t}
\newcommand{\history}{h}
\newcommand{\future}{f}
\newcommand{\dialog}{g}
\newcommand{\numTurns}{n}
\newcommand{\sent}{s}

\newcommand{\myvec}{v}

\newcommand{\outvec}{\myvec_{out}}
\newcommand{\score}[1]{Score(#1)}

\newcommand{\arbmat}{W}
\newcommand{\arbvec}{b}

\newcommand{\origBERT}{{IREP}\textsubscript{BERT}\xspace}
\newcommand{\rankBERT}{{RANK}\textsubscript{BERT}\xspace}
\newcommand{\rankBERTArg}[1]{\rankBERT(#1)\xspace}

\newcommand{\retList}{L}
\newcommand{\rank}[2]{rank_{#1}(#2)}

\newcommand{\prob}{p}
\newcommand{\ilmprob}{\prob}
\newcommand{\inducedprob}[3]{\ensuremath{#1_{#2}(#3)}}
\newcommand{\dirLM}[2]{\inducedprob{\ilmprob^{Dir}}{#1}{#2}}
\newcommand{\mleProb}[2]{\inducedprob{\ilmprob^{MLE}}{#1}{#2}}
\newcommand{\mixProb}[2]{\inducedprob{\ilmprob^{MIX}}{#1}{#2}}

\newcommand{\arbText}{x}
\newcommand{\condP}[2]{\prob(#1\vert#2)}
\newcommand{\docSet}{{\mathcal D}}
\newcommand{\sentSet}{{\mathcal S}}

%% file: category.tex

\ccsdesc[500]{Information systems}
\ccsdesc[500]{Information systems~Information retrieval}
\ccsdesc[500]{Information systems~Retrieval tasks and goals}

\keywords{dialogue retrieval; sentence retrieval}

%% file: abstract.tex
\begin{abstract}
We address the task of sentence retrieval for open-ended dialogues. The goal is to retrieve sentences from a document corpus that contain information useful for generating the next turn in a given dialogue.  Prior work on dialogue-based retrieval focused on specific types of dialogues: either conversational QA or conversational search. To address a broader scope of this task where any type of dialogue can be used, we constructed a dataset that includes open-ended dialogues from Reddit, candidate sentences from Wikipedia for each dialogue and human annotations for the sentences. We report the performance of several retrieval baselines, including neural retrieval models, over the dataset. To adapt neural models to the types of dialogues in the dataset, we explored an approach to induce a large-scale weakly supervised training data from Reddit. Using this training set significantly improved the performance over training on the MS MARCO dataset.
\end{abstract}

%% file: intro.tex
\section{Introduction}
\label{sec:intro}

\begin{figure}[t]
    \makebox[1 \columnwidth][c]{
    \includegraphics[width=0.8\columnwidth, height=6cm]{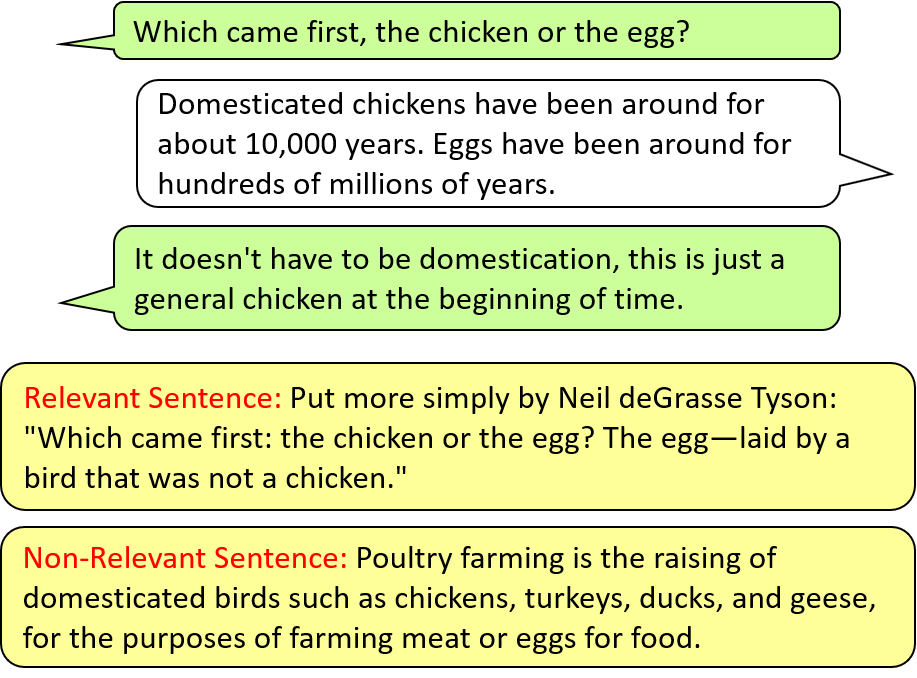}
    }
    \caption{Test example of a dialogue created from Reddit with sentences judged as relevant and non-relevant by human annotators.}
    \label{fig:conv-example}
\end{figure}

%The last few years have been 
There has been a rapid increase in the last few years
in research of tasks related to dialogue (conversational) systems \cite{Gao2018a,WizardOfWiki,Reddy+al:19a,Dalton+al:20a,Gao+al:22a,Zamani+al:22a}. 
Prominent examples include response generation \cite{WizardOfWiki,TopicalChat,Zhu2018a,qin-etal-2019-conversing} or response selection \cite{Yan+al:16a,QuAC,Qu+al:19a,Stamatis+al:19a,Ren+al:18a,Yan+Zhao:18a,zhang2018modeling,Huang+al:19a} with respect to the last turn in the dialogue, conversational question answering \cite{Elgohary+al:18a,Zhu2018a,Lin+al:20b} and conversational retrieval (of passages) \cite{Radlinski+Craswell:17a,Dalton+al:20a,Gao+al:22a,Zamani+al:22a}. 

In this paper we focus on open-ended dialogues: two parties converse in turns on any number of topics with no restrictions to the topic shifts and type of discussion on each topic. In addition, the dialogue is not grounded to a specific document, in contrast to the setting used in some previous work (e.g., \cite{Ma+al:20a}).
The task we address is retrieving sentences from some document corpus that contain information useful for generating (either automatically or by humans) the next turn in the dialogue. We note that the dialogue turns can be questions, queries, arguments, statements, etc.  

Existing dialogue/conversational datasets are not well suited for evaluation of the task we pursue; we discuss this point in length in Section \ref{sec:rel}.  Hence, we developed a novel dataset reported here. The dataset includes 846 dialogues created from Reddit threads. For each dialogue, $50$ sentences were retrieved from Wikipedia using an unsupervised initial retrieval method. These sentences were judged by crowd workers for relevance, that is, whether they contained information useful for generating the next turn in the dialogue. Figure \ref{fig:conv-example} depicts one such dialogue, with two sentences annotated by the raters, one as relevant and one as non-relevant. The dataset is available at \url{https://github.com/SIGIR-2022/A-Dataset-for-Sentence-Retrieval-for-Open-Ended-Dialogues.git}. 

Neural-based retrieval methods require lots of training data --- whether for learning from scratch or for fine tuning a pre-trained model. Hence, we used a weak supervision approach to induce pseudo relevance labels for a few sentences for each of $\sim$73,000 additional dialogues. To this end, we fused rankings induced by several methods over an initial %retrieved\itay{why retrieved? we didn't use any method for retrieval; they're just the sentences from the pointed document/section} 
sentence list. These methods are either unsupervised or are based on distant supervision. For example, we used a BERT-based model \cite{BERT} trained for query-based passage retrieval on the MS MARCO dataset \cite{nogueira2019passage}. 

We report retrieval performance over the dataset for several methods. 
% Some of these utilize the BERT pre-trained contextualized language model. 
The evaluation demonstrates the clear merits of using the pseudo relevance labels induced using the weak supervision to fine tune BERT-based retrieval methods.
%%-OK: I think it is not a major point to emphasize in the intro
%It also shows performance differences between two dialogue types - those with related Wikipedia articles in Reddit metadata, and those without such metadata.  

Our contributions can be summarized as follows:
\begin{itemize}

\item A dataset for open domain dialogue systems with data labeled for the task of retrieving sentences that contain useful information for response generation.

\item A procedure to induce a large scale weakly supervised training data using the metadata of Reddit dialogues.

\end{itemize}

%% file: rel.tex
\section{Related Work}
\label{sec:rel}

There are two main lines of work related to ours. The first is on devising datasets for conversation systems. The second line is on using weak supervision for retrieval. 

\subsection{Conversational Datasets}
\label{ssec:rel_datasets}

Existing conversational datasets were built for two main tasks. The first is to compose the next turn in a dialogue, either via generative language models or by retrieving full responses from an index. Therefore, related datasets \cite{Yan+al:16a, qu2018analyzing, WizardOfWiki, TopicalChat, wang2013dataset, lowe2015ubuntu, dailydialog, MovieDialogue, zhang2018personalizing, moghe-etal-2018-towards, zhou-etal-2018-dataset, akasaki-kaji-2019-conversation, DSTC7}
%%-OK:
%evaluates
serve to evaluate
the offered responses compared to gold responses, not the retrieval of relevant information for composing such responses. The second task focuses on conversational passage retrieval and question answering (QA), where information needs are conveyed by a user to a search engine or a question answering system via a series of queries/questions that should be considered a single session. Prominent conversational QA datasets include CoQA \cite{Reddy+al:19a}, DoQA \cite{DoQA} and QuAC \cite{QuAC}. In these datasets, all evaluation sessions are grounded to a single passage or section that the participants are allowed to ask and answer about. In contrast, we address dialogue-based sentence retrieval from a document corpus.  

The only conversational passage retrieval dataset we are familiar with is from TREC's CAsT tracks \cite{Dalton+al:20a,Dalton+al:21a}. However, CAsT's queries reflect explicit intents, while we are also interested in more open dialogues where the information needs can be in the form of implicit intents, as shown for example in Figure \ref{fig:conv-example}.
In these datasets, the user conducts a query session on a specific single topic. The queries in the session may co-refer and reflect prior queries in the session. However, in most of these datasets, the returned search results are not viewed as part of the dialogue.
%and the queries only co-refer previous user queries, not search results.  
Finally, in both conversational passage retrieval and conversational QA datasets, there is a user asking questions or queries that reflect explicit intents with information needs, as opposed to natural dialogues where intents may be only implicitly represented, e.g., in affirmative statements.
%, or not at all; e.g., in affirmative statements or social chat.

To sum, existing conversational datasets do not combine natural human-human conversations with relevance annotations for sentences retrieved from a large document corpus. We therefore constructed  such a dataset and present it in Section \ref{sec:dataset}.

\subsection{Weak Supervision for Retrieval}
\label{ssec:weak_sup_rel}
%In earlier work on neural retrieval, pseudo relevance labels attained by using a classical retrieval method were used for weak supervision \cite{Dehghani+al:17a,Zamani+Croft:18a}. We fuse rankings induced by classical methods and modern ones (based on BERT) to induce pseudo relevance labels.

A standard approach for using neural models in IR (and text-based tasks in general) , specifically dense retrieval models \cite{Lin+al:21a}, is to first pre-train the neural model, either on a different target task but with a lot of training data, such as masked language modeling, or on a similar task as the end-goal, but on a different domain. Then, the pre-trained model is fine-tuned on training data for the target task or domain \cite{BERT, nogueira2019passage,li2020parade}. For example, \citeauthor{yilmaz2019cross} \cite{yilmaz2019cross} fine-tuned a sentence retrieval model by first training a retrieval model on the MS MARCO retrieval task, and then fine-tuning it on a Microblog retrieval dataset.

Fine-tuning of retrieval models requires relevance labels for training examples in a target task. These are sometimes scarce or unavailable. One approach to circumvent this is to automatically generate labels and train a weakly supervised model on these annotations.  \citeauthor{Wu+al:18a} \cite{Wu+al:18a} trained a response selection model for conversations using a weak signal induced from a matching probability offered by a seq2seq model trained on human conversations. \citeauthor{li2019learning} \cite{li2019learning} used weak annotators based on search logs to train a Search Ads matching model. They automatically selected pseudo negative examples by optimizing the distance between a pseudo negative example from a retrieved pool and a given positive example. \citeauthor{Dehghani+al:17a} \cite{Dehghani+al:17a} used Okapi BM25 rankings to induce pseudo relevance labels so as to train a neural model ranker. \citeauthor{Zamani+Croft:18a} \cite{Zamani+Croft:18a} then provided a theoretical analysis for the merits of using such weak supervision for ad hoc retrieval.
Weak supervision was also used for other related retrieval tasks; for example, expanding the last dialogue \post for passage retrieval \cite{Voskarides+al:20a} and for query performance prediction \cite{Zamani+al:18a}.

%. For example,  \citeauthor{Voskarides+al:20a} \cite{Voskarides+al:20a} used weak supervision to expand the last \post in a dialogue. The expanded \post was used for passage retrieval. \citeauthor{Zamani+al:18a} \cite{Zamani+al:18a} used weak supervision for query performance prediction.

We follow the weak supervision paradigm in our model training, with a novel weak Reddit annotator for retrieval in a dialogue context.

% \begin{itemize}
%   \item \cite{Wu+al:18a}  Ran initial retrieval; then used the probabilities assigned by a seq-to-seq model (trained on human conversation data) to induce a weak signal for training
%   \item \cite{Li+al:19a} focused on automatically selecting pseudo negative examples during the training of a model and using its already learned metric. (specifically, they tried to optimize the distance between a pseudo negative example from a retrieved pool and a given positive example
%   \end{itemize}

%% file: dataset.tex
\section{A Dataset for Open Dialogue Retrieval}
\label{sec:dataset}

%Our task in this paper is information retrieval for open-ended dialogue context. A proper evaluation setup for this task should include representative dialogues and candidate retrieved items with relevance ranking for these contexts. 
%However, publicly available datasets that link between a textual source and a \post in a dialogue were mainly created with respect to text generation tasks and not information retrieval tasks \cite{moghe-etal-2018-towards, WizardOfWiki, TopicalChat, zhou-etal-2018-dataset, akasaki-kaji-2019-conversation, DSTC7}.
%Such datasets contain annotations to the specific sentence, passage or document that was used by a human to generate \posts. But, they do not include indication whether other retrieval candidates could be helpful for generating the same or a different yet still relevant \post within the same dialogue context.
%In addition, these datasets usually constrain the knowledge selection for grounding each response in a given dialogue to a pre-collected small set of related passages, usually originating from a single document such as a Wikipedia article. This is opposed to our aim at retrieving relevant content from a large collection of documents for any open-ended dialogue.

As described in Section \ref{ssec:rel_datasets}, we are not aware of a
dataset that can be used to evaluate
%information
retrieval in an open-ended dialogue
context.  We therefore constructed such a dataset.
%specifically for evaluating the task of information retrieval in open-ended dialogue context. 
To this end, we %extracted 
used dialogues from Reddit,
%%-OK: also for the train data
%as our test set
due to its large variety of topics and styles. We then retrieved candidate Wikipedia sentences
%%-OK: the retrieval is for the dialogue
%for \posts in
for %the extracted 
these dialogues. 
Each candidate was judged for relevance
%rated
by crowd-source human annotators; it was marked as relevant if it contained information useful for generating the next turn in the dialogue.
%for relevancy to the respective conversation context by answering whether the candidate could be helpful in writing the next \post in the dialogue. 
%In the training-set part, to achieve annotation at large-scale, we propose an automatic labeling algorithm within a weakly supervised training scheme.
We next detail the way we collected dialogues from Reddit and the manual annotation process and construction of the
%test set.
dataset.

%Similarly to the above test set, we wanted to construct a training set tailored for the dialogue retrieval task. Prior work show that large train training sets significantly boost the performance of modern neural networks [?]. However, while our the manual annotation we employed for the test-set construction is of high-quality, it is also slow and expensive. We therefore study an alternative approach using automatic labeling algorithm that would generate a large scale training set for sentence retrieval in dialogue context within a weakly supervised training scheme. 
%To train modern retrieval models, which use data hungry neural networks, a large-scale training set is required [?]. The human annotation process we employed for generating our test set is expensive and time consuming, and while useful for generating high-quality evaluation sets, it is prohibitively slow for collecting large training data. Therefore, in this work we explore an algorithmic approach for generating a large-scale training set for sentence retrieval in dialogue context. We next describe our training set collection and the pseudo labeling of each training example.

%We next detail the way we collected dialogues from Reddit, the manual rating process and construction of the test set, and the automatic training-set labeling.

\subsection{The Reddit Collection}
\label{ssec:collection}

Reddit is a multi-party discussion platform, where users add \posts in a conversation that can be replied to by several users. Each discussion is therefore a tree structure of \posts, and a path from the initial \post (the tree's root) to a \post without followup replies (a leaf) is called a \emph{thread}. 
We collected $263,986,046$ conversation threads from Reddit after preprocessing the submission and comment files provided by the Pushift.io archives\footnote{\url{https://files.pushshift.io/reddit/}}. All threads are in English, none of them belongs to topics (called \emph{subreddits} in Reddit) on offensive content, and they do not contain missing \posts, i.e., \posts that have been removed and their text is not displayed. We refer to the collection of these threads as the \emph{Reddit collection}.

%A single thread in Reddit represents a multilogue. To distill dialogue-like conversations from threads, we considered only threads in which the user that initiated the discussion, called here the \emph{\asker}, is the author of each odd turn of the thread; all other users in the thread, who author the even turns, are considered the \emph{\answerers}. These threads are thus interleaving \posts between the \asker and \answArt \answerer. We refer to these threads as \emph{Reddit dialogues} and the whole collection as the \emph{Reddit dialogue collection}.

%Finally, we kept only dialogues with four \posts or more, so that the retrieval task would be exposed to enough dialogue context. 
%These dialogues constitute our \emph{Reddit dialogue collection}.
%We filtered out Reddit dialogues of three turns or less. This way, both the \asker and the \answerers participate at least twice in the dialogue, offering a context of at least a two-turn dialogue history for the later \posts. The remaining dialogues constitute our \emph{Reddit dialogue collection}.

We enriched the context of each thread by prepending to the first \post: (a) the name of the subreddit the discussion was issued under, and (b) the title of the discussion (if provided by the user who initiated the thread).  %In each dialogue, we randomly picked a specific \answerer \post as the \emph{\target}. Then, each dialogue essentially ends with the \post preceding the \target.
%This \post serves as an example for a valid \post to be generated based on preceding \posts in the dialogue and information in sentences deemed relevant.
%We only picked \targets with at least 3 predecessor \posts to have a meaningful dialogue context. 
We split the threads by dates: test set candidates (discussed in Section \ref{ssec:test set}) were limited to dates between Feb-2011 and Dec-2013, and training set candidates (discussed in Section \ref{sec:distant_supervision}) were limited to dates between Jan-2014 and Aug-2019.

Some \posts in Reddit offer one or more links to Web pages that the
author considers related to the written text. We only considered links
to specific sections on Wikipedia pages that are found in our
downloaded version of Wikipedia; we refer to these \posts as
\textit{Wikipedia grounded \posts}.

\subsection{Test Set Construction}
\label{ssec:test set}

%\oren{It is unclear as we haven't defined a test set split}\idan{The split is presented in the prev subsection}
%As candidates for the test set,
To construct a test set, we randomly sampled threads from
the test set part of our Reddit collection which was described in Section \ref{ssec:collection}. A single thread in Reddit represents a multilogue. Hence, to distill dialogue-like conversations from threads, we considered only threads in which the user who initiated the discussion, called here the \emph{\asker}, is the author of each odd turn of the thread. All other users in the thread, who author the even turns, are considered the \emph{\answerers}. These threads are thus interleaving \posts between the \asker and the \answerers.
Threads with less than 4 \posts were discarded, since we want a meaningful dialogue context.
We refer to
%these
the remaining
threads as \emph{Reddit dialogues}.
In each candidate
dialogue,
%we selected the last \answerer \post
we refer to the last \answerer \post
%we randomly selected an \answerer \post,
%call the \emph{\target}, with at least 3 predecessor \posts
as the \emph{\target}.
%(to have a meaningful dialogue context)\footnote{If no such \post exist, the dialogue was discarded.}.
A \emph{test dialogue} was constructed by trimming the
candidate dialogue to include \posts up to, but excluding the \emph{\target}.
%We randomly picked dialogues, where leaves whose author is the \emph{\asker} were removed. The resulting leaves (where their author is necessarily a \answerer) are denoted as \emph{\targets}.  We only picked \targets with at least 3 predecessor \posts to have a meaningful dialogue context. These threads essentially ends with the \post preceding the \target written by the \answerer.

%%-OK: we don't annotate the dialogue
%To annotate a candidate dialogue,
We used an initial
%ranker
sentence retrieval method (henceforth also referred to as an initial ranker), described in Section \ref{sec:initial_ranker}, to retrieve $50$ candidate sentences from Wikipedia for
%to
%the
each test dialogue. The retrieval method utilizes all \posts in the test dialogue.
%%-OK: I believe this is discussed later
%context constituting all \posts in the dialogue up to but excluding the \target.
We then recruited master\footnote{https://www.mturk.com/worker/help\#what\_is\_master\_worker}
%\hagai{Is that a known title of some of the raters? If not, we should explain (here or in the appendix) how we chose them}\itay{Yes, it's a known qualification on MTurk.}
workers from Amazon Mechanical Turk to judge the relevance of each retrieved sentence.
%, where relevance refers to whether the candidate sentence may be helpful in writing the next \post in the dialogue.
%%-OK: defined above.
%A candidate is considered \textit{relevant} if it contains sufficient information for writing a \post that can be a natural continuation to the dialogue. Otherwise, it is considered \textit{non relevant}.
Our instructions, presented in Figure \ref{fig:mturk_guidelines},
%(\ref{fig:mturk_guidelines})
were to mark a sentence as relevant to a dialogue if it contained information useful for generating the next \post in the dialogue that would be a natural continuation. We provided the annotators with the dialogue \posts, including its topic and title, and the Wikipedia sentences including their page's title.
%; recall that the dialogue ends with the \post just before the \target.
%Annotators were only shown the \posts in the test dialogue and the dialogue's title. 
The definition of sentence relevance ---
as conveyed to the annotators --- does not indicate the correctness or soundness of a \post that might be generated based on information in
the sentence.

Every Mechanical Turk worker had to annotate 10 dialogues and 5 retrieved sentences for each dialogue.
%For each dialogue, we provided its title and \posts.
%in the dialogue up to but excluding the \target.
We truncated the \posts and sentences to the first 30 words for ease of use; the workers had access to the whole text by hovering the cursor over it.
%The annotation guidelines are presented in Figure \ref{fig:mturk_guidelines}.
At most three workers
%were required to
%assign a sentence candidate as relevant or non-relevant
judged a sentence for relevance. We started with a majority vote between two workers, and if no majority agreement was achieved, the third vote was then used. %About $7\%$ of the sentences among Wikipedia grounded dialogues required a third round, and about $4\%$ among the non-grounded dialogues. %\hagai{I don't understand the above sentence. 
%\hagai{Can we show stats of relevancy among the candidates that needed to have 3 ratings?} \itay{I mentioned above estimated percentages; I have a mess with the results}\hagai{Estimate based on what?}
%\hagai{Maybe show one example of disagreement?}
%As mentioned above, the Wikipedia grounded dialogues have a tendency to yield more relevant candidates rather than non-grounded dialogues. 
We used honeypots to identify and disqualify raters that seemed to not follow our guidelines. % \hagai{Add example in the appendix?}
An example dialogue from our test set with one
%candidate
sentence judged as relevant and one
%rated
as non-relevant is shown in Figure \ref{fig:conv-example}.

Since we show candidate test set dialogues to humans, we filtered out
test dialogues whose content includes one or more words from a
public list of dirty
words\footnote{https://github.com/LDNOOBW/List-of-Dirty-Naughty-Obscene-and-Otherwise-Bad-Words/blob/master/en}.
We also expanded this list by concatenating phrases in the list to
detect hashtags and titles consisting dirty words without spaces.  In
addition, we filtered out dialogues whose \posts contain URL
references (with the exception of the Wikipedia grounded target
\posts, which are not shown to the raters), and those with \posts
%that
%are
shorter than 5 tokens and longer than 70 tokens, as they resulted in difficult relevance annotations.
%%-OK: we don't rate the dialogues
%were
%found difficult to rate.
We only kept test dialogues for which at least one sentence was judged as relevant.

We noticed that sentences retrieved for dialogues whose \target is a Wikipedia grounded \post (i.e., it includes a reference to a Wikipedia section)
%have higher tendency to receive relevance ratings for candidate retrieved sentences
were somewhat more likely to be judged as relevant than those retrieved for dialogues whose \target was not Wikipedia grounded. See Table \ref{tab:test_set_rel_cands_stats} for details.
%%-OK: not clear what adequate is
%This could indicate that the retrieval task would be more adequate for such dialogues. 
Following, we denoted dialogues with Wikipedia grounded \targets as \textit{Wikipedia grounded dialogues}, and balanced our test set between
%Wikipedia grounded dialogues
such dialogues
and ungrounded dialogues.
% If the \target is Wikipedia grounded, i.e. it includes a reference to a Wikipedia page,  we call the whole dialogue a \textit{Wikipedia grounded dialogue}.
%
%To make the best use of the test set for our IR task,
% We only kept candidate dialogues for which
%that were annotated with at least one relevant
% at least one sentence was judged as relevant.
%We thus continued the annotation process until we reached a sufficient number of annotated dialogues.
%This process resulted in
 As a result, the test set includes $846$ dialogues with relevance-judged sentences: $400$ ungrounded dialogues and $446$ Wikipedia grounded dialogues. 
%In our work, we split this data into 50 random splits of validation and test, while the number of non-grounded and grounded dialogues is equal in each split.

We found that using the thread's subreddit name and title (if exist) is highly important for the initial retrieval method
described in Section \ref{sec:initial_ranker}, since many times the first turn might be missing or very redundant since the user provided the needed information in the thread title. For example, $202$ and $155$ dialogues out of $446$ Wikipedia grounded dialogues and $400$ ungrounded dialogues, have an empty first \post.
%\hagai{Are all the targets in this dataset are leaves? Is this differ in train/test set? I remember we used future post for pseudo labeling, i.e., the targets are not necessarily the leaves. Maybe I miss that, but I don't see where we explain that we don't take the entire thread as a conversation.}\itay{We only used future posts for the pseudo-labeling process; none of them are target posts.}\idan{we dont say anywhere that we selected leaves as target posts, only those with sufficient context - enough dialogue history}
%\input{summaryStats}
%\input{dataset_stats}

\begin{table}[t]{
    \caption{\label{tab:test_set_rel_cands_stats} Test set statistics (average, median and standard deviation) of the number of relevant sentences and the rank of the first relevant sentence (the highest rank is 1) retrieved by the initial ranker per dialogue.}
\small
\input{tables/test_set_rel_cands_stats}}
\end{table}

\begin{figure}[t]
    \makebox[1 \columnwidth][c]{
    \includegraphics[width=0.9\columnwidth, height=5cm]{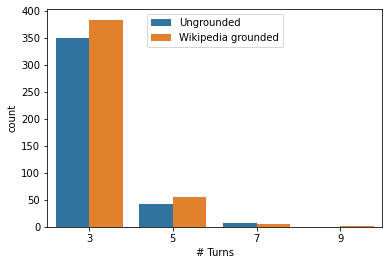}
    }
    \caption{Dialogue count breakdown by the number of turns in the dialogue history (excluding the target \post).}
    \label{fig:num_turns_vs_type}
\end{figure}

\begin{figure}[t]
    \makebox[1 \columnwidth][c]{
    \includegraphics[width=0.9\columnwidth, height=5cm]{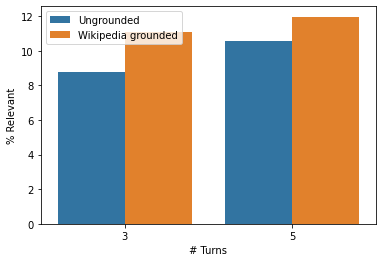}
    }
    \caption{Percentage of relevant candidate sentences for dialogues with a history of $3$ or $5$ \posts.}
    \label{fig:num_turns_vs_pos}
\end{figure}

Figure \ref{fig:num_turns_vs_type} shows that most of the test dialogues have a history of 3 turns.
%The dialogues with more turns seems to have more relevant sentences, in average, in their initial list,
Dialogues with a history of 5 turns have more relevant sentences in the initially retrieved list than dialogues with 3 turns in the history,
as shown in Figure \ref{fig:num_turns_vs_pos}. Dialogues with with history longer than 5 \posts are not considered in Figure \ref{fig:num_turns_vs_pos} since the test set includes only a few of them (see Figure \ref{fig:num_turns_vs_type}).

% It is shown in Figure \ref{fig:percentage_rel_vs_num_terms} that for \emph{Wikipedia grounded} dialogues there is a trend, that as the number of terms in the last turn increases, the percentage of candidate sentences retrieved in the initial list is higher. This observation does not hold for \emph{Ungrounded} dialogues.

\input{initial_ranker}

%% file: tables/test_set_rel_cands_stats.tex
\begin{tabular}{|c|c|c|c|c|c|c|} 
\hline
Dialogue type & \multicolumn{3}{c|}{Ungrounded} & \multicolumn{3}{c|}{Wikipedia Grounded} \\
\cline{2-7}
& Avg & Med & Std & Avg & Med & Std \\
\specialrule{.2em}{.1em}{.1em}
%\hline
%\hline
\# Relevant & 4.43 & 3 & 4.45 & 5.68 & 4 & 5.09 \\
\hline
Rank of 1st Relevant & 11.3 & 6 & 12.5 & 9.395 & 4 & 12.14 \\
\hline
\end{tabular}

%% file: initial_ranker.tex
\subsection{Initial Sentence Retrieval}
\label{sec:initial_ranker}
% Let $\dialog$ denote a dialogue composed of the \posts: $\ut_1,
% \ldots, \ut_{\numTurns}, \ut_{\numTurns+1}$; $\ut_1$ is the first
% \post and $\ut_{\numTurns+1}$ is the \target which is not used
% for retrieval; $\ut_{\numTurns}$ is the last turn.

Let $\dialog$ denote a dialogue composed of the \posts: $\ut_1, \ldots, \ut_{\numTurns}$; $\ut_1$ is the first \post and $\ut_{\numTurns}$ is the last turn. 
To retrieve sentences for $\dialog$, we follow common practice in work on passage retrieval \cite{Corrada-Emmanuel+al:03a,Murdock:07a} and first retrieve documents. %We then rank the sentences in these documents.\itay{at first glance, it appears we ranked the sentences w.r.t. each document separately.}
We then rank their sentences.
The retrieval methods we present are unsupervised; they utilize unigram language models \cite{Croft+Lafferty:03a}. 

We start by describing notation. We use $\mleProb{\arbText}{\arbTerm}$ to denote the maximum likelihood estimate (MLE) of term $\arbTerm$ with respect to the text, or text collection, $\arbText$. $\dirLM{\arbText}{\arbTerm}$ is the probability assigned to $\arbTerm$ by a Dirichlet smoothed unigram language model induced from $\arbText$ \cite{Zhai+Lafferty:01a}. We compare two (unigram) language models, $\theta_1$ and $\theta_2$, using the cross entropy: $$\ce{\condP{\cdot}{\theta_1}}{\condP{\cdot}{\theta_2}} \definedas -\sum_{\arbTerm} \condP{\arbTerm}{\theta_1}\log \condP{\arbTerm}{\theta_2}.$$ 

%We implemented an initial ranker which utilizes unigram language models (LMs) and induces contextual information from the dialogue. We applied common practice in IR and used the following three-step procedure: document ranking, sentence ranking and combining their scores. Let ${t_{1},..., t_{n}, t_{n+1}}$ be the turns in dialogue ${q}$ of length ${n+1}$, where ${t_{n+1}}$ is the target \post, which is not used in the retrieval steps.

\subsubsection{Retrieval Method}
\label{sec:docRank}
For document retrieval, we represent the dialogue $\dialog$ as a linear mixture of language models induced from its \posts:
\begin{equation}
  \label{eq:dcoRepDiag}
  \condP{\arbTerm}{\dialog^{Doc}} \definedas (1-\beta)\mleProb{\ut_1}{\arbTerm} + \frac{\beta}{\numTurns-1} \sum_{i=2}^{\numTurns} \mleProb{\ut_{i}}{\arbTerm};
  \end{equation}
$\beta$ is a free parameter. Since the first \post, $\ut_1$, also contains the dialogue title and the subreddit, we assign it a specific weight.
The document corpus is Wikipedia. A Wikipedia document, $\doc$, is scored with respect to $\dialog$ using:
$$Score(\doc;\dialog)
\definedas - \ce{\condP{\cdot}{\dialog^{Doc}}}{\dirLM{\doc}{\cdot}}.$$
The outcome of the document retrieval step is $\docSet$: the set of top-$k$ retrieved documents.
%\begin{equation}
%\label{eqn:CE_docs}
%Score(d;q) \definedas - \ce{\condP{\cdot}{\dialog^{Doc}}}{\condP{\cdot}{\doc}}.
%\end{equation}

The next step is to rank the sentences $\sent$ in $\sentSet$: the set
of all sentences of documents in $\docSet$.
%We use $\doc_{\sent}$ to denote the ambient document of $\sent$.
For sentence retrieval, we represent
the dialogue using a mixture model again. But, in contrast to Equation \ref{eq:dcoRepDiag}, the emphasis now is on the last \post, $\ut_n$: 
\begin{equation}
\label{eq:sentRepDiag}
\condP{\arbTerm}{\dialog^{Sent}} \definedas (1-\beta)\mleProb{\ut_{\numTurns}}{\arbTerm} + {\beta} \sum_{i=1}^{\numTurns-1} \alpha_i \mleProb{\ut_{i}}{\arbTerm};
\end{equation}
%where
${\alpha_i}\definedas{\frac{{\delta}e^{-{\delta \hspace{0.2mm}\abs{T-i}}}}{\sum_{j\in\indSet}{{\delta}e^{-{\delta\hspace{0.2mm}\abs{T-j}}}}}}$; $T=n-1$, $\indSet = \{1, ..., n-1\}$ and $\delta$ is a free parameter (cf., time-based language models \cite{Li+Croft:03a}). The rationale is that the next \post to be generated for the dialog should be, to some extent, a response to the last \post $\ut_{\numTurns}$. The preceding \posts serve as the dialogue context and their induced language models are weighed using an exponential decay function.  The direct retrieval score of $\sent$ is defined as:
$$Score(\sent;\dialog) \definedas - \ce{\condP{\cdot}{\dialog^{Sent}}}{\dirLM{\sent}{\cdot}}.$$

Finally, following common practice in work on sentence retrieval \cite{Murdock:07a}, we integrate the direct retrieval score of $\sent$ with the retrieval score of its ambient document $\doc_{\sent}$:
\begin{equation}
\label{eq:final_score}
FinalScore(\sent;\dialog) \definedas (1-{\gamma}){Score'(\doc_{\sent};\dialog)} +{\gamma}Score'(\sent;\dialog),
\end{equation}
where $Score'(\doc_{\sent};\dialog)$ and $Score'(\sent;\dialog)$ are the min-max normalized $Score(\doc_{\sent};\dialog)$ and $Score(\sent;\dialog)$ with respect to $\docSet$ and $\sentSet$, respectively; $\gamma$ is a free parameter.
%\end{equation}

%Table \ref{tab:candidate_stats} shows that candidate sentences retrieved for \emph{Wikipedia grounded} dialogues, contain more words on average and also retrieved with a bit more \initRank score, compared to sentences retrieved for \emph{Ungrounded} dialogues. It is also shown that the \initRank score is higher for candidate sentences that are marked as relevant, compared to irrelevant candidate sentences.

Table \ref{tab:candidate_stats} shows that candidate sentences retrieved for \emph{Wikipedia grounded} dialogues contain more words on average than sentences retrieved for ungrounded dialogues.
Relevant (non-relevant) sentences retrieved for grounded dialogues receive higher retrieval scores by the \initRank than relevant (non-relevant) sentences retrieved for ungrounded dialogues.  For both grounded and ungrounded dialogues, the average retrieval score for a relevant sentence is higher than that for a non-relevant sentence.

\begin{table}[t]{
    \caption{\label{tab:candidate_stats} Number of words and the \initRank retrieval score for candidate relevant and non-relevant sentences in the initially retrieved list.}
\small
\input{tables/candidates_stats}}
\end{table}

\begin{figure}[t]
    \makebox[1 \columnwidth][c]{
    \includegraphics[width=0.9\columnwidth, height=5cm]{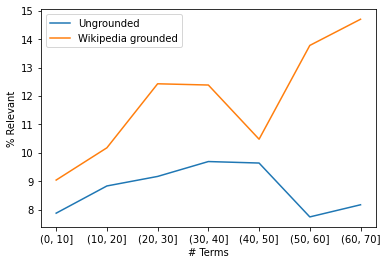}
    }
    \caption{The effect of the number of terms in the last \post ($t_n$) in the dialogue on the percentage of relevant sentences in the initially retrieved list.}
    \label{fig:percentage_rel_vs_num_terms}
\end{figure}

Figure \ref{fig:percentage_rel_vs_num_terms} presents the percentage of relevant sentences in the initially retrieved list as a function of the number of terms in the last turn of the dialogue ($t_n$). We see an overall increasing trend with the main exception being very long turns for ungrounded dialogues. The upward trend can be explained as follows. The sentence ranker puts emphasis on the last turn ($t_n$). With more information in this turn, the higher the likelihood to retrieve sentences that contain useful information to produce the next response. These are the sentences that are deemed relevant.

%% file: tables/candidates_stats.tex
\begin{tabular}{|c|c|c|c|c|c|c|} 
\hline

Dialogue type & Label & \# Words & \initRank score \\
\specialrule{.2em}{.1em}{.1em}
%Ungrounded & Non-relevant	& 29.999725696730305 &	0.530807113283704 \\
Ungrounded & Non-relevant	& 30 &	0.53 \\
\hline
%Ungrounded & Relevant	 &	28.834650112866818 &	0.5662680989103169 \\
Ungrounded & Relevant	 &	28.83 &	0.57 \\
\specialrule{.2em}{.1em}{.1em}
%Wikipedia grounded & Non-relevant & 31.699721730331394 &		0.5473008986760345 \\
Wikipedia grounded & Non-relevant & 31.7 &		0.55 \\
\hline
%Wikipedia grounded & Relevant &	31.365285996055228 &		0.5918497196688421 \\
Wikipedia grounded & Relevant &	31.37 &		0.59 \\
\hline
\end{tabular}

%% file: models.tex
\section{Sentence Retrieval Models for Open-Ended Dialogues}
\label{sec:models}
To provide some sentence retrieval performance results on our proposed
dataset, in addition to those of the initial retrieval method
described in Section \ref{sec:initial_ranker}, we used a few
methods described in Section \ref{ssec:eval_models}. Some of these (e.g., neural-based) require a (large) training set. 
%we evaluated several IR models, both token-matching
%models and recent neural models.  To train neural models, a labeled
%training set for the task at hand is required.
However, generating a large scale manually annotated training set is a
laborious effort. Therefore, we instead propose in Section \ref{sec:distant_supervision} a weakly supervised
method for automatic labeling.

%In Section \ref{ssec:eval_models} we present the dialogue-based sentence retrieval methods we experimented with. In Section \ref{sec:distant_supervision} we describe our weak supervision approach.
%We next detail the models we assessed in our experiments
%and our weakly supervised training set.

\subsection{Retrieval Methods}
\label{ssec:eval_models}
%Suppose that two parties are holding a dialogue which is an open-ended conversation. That is, they discuss a topic, or multiple topics, may express information needs (e.g., via questions), try to satisfy these needs, can make arguments and counter arguments, etc. Let $\ut_{\numTurns}$ denote the current (last) \post. Herein, we refer to a \post and the text it contains interchangeably. Our goal is to retrieve ``relevant'' sentences from some document corpus $\corpus$, where relevance means that the sentence contains information that can be used to generate the next \post, $\ut_{\numTurns+1}$.

%In contrast to the ad hoc retrieval task, where the information need is explicitly expressed via a query, in open-ended dialogue we have no explicit expression.
%One proxy to the presumed information need is the current \post $\ut_{\numTurns}$: in many cases, the next \post to be generated, $\ut_{\numTurns+1}$, is a direct response to $\ut_{\numTurns}$. Accordingly, we set the input query to be $\ut_{\numTurns}$, and match it against a retrieved candidate sentence $\sent$. 

%In this paper, we address the retrieval task for open-ended dialogue via a two-stage ranking approach: an initial ranker (Section \ref{sec:initial_ranker}) retrieves $k$ candidate sentences, which are then reranked by each of the following methods:

%For each $\query$, the top-$k$ retrieved sentences were reranked using the lexical retrieval methods:
We employ a two step sentence retrieval approach. First, we apply the
initial sentence retrieval method described in Section
%\ref{ssec:eval_models}
\ref{sec:initial_ranker}. Then, we re-rank the top-$k$ retrieved
sentences using one of the methods proposed below.  In contrast to the
ad hoc retrieval task, where the information need is explicitly
expressed via a query, in open-ended dialogues we have no explicit
expression.  One proxy to the presumed information need is the last
\post $\ut_{\numTurns}$: in many cases, as noted above, the next \post to be
generated is a response to
$\ut_{\numTurns}$. Accordingly, herein we use $\ut_{\numTurns}$ as a
query for (re-)ranking the sentences in the initially retrieved
list. Utilizing the dialogue context in the retrieval methods
described below, as was the case for the initial sentence retrieval
method, is left for future work.

\myparagraph{LM} We score sentence $\sent$ using $-\ce{\mleProb{\ut_{\numTurns}}{\cdot}}{\dirLM{\sent}{\cdot}}$.

\myparagraph{Okapi} We assign $\sent$ its BM25 retrieval score with respect to $\ut_{\numTurns}$ \cite{Robertson+al:94}.

%We also applied two additional retrieval methods that are based on contextualized language model (BERT) and estimate the match between $\query$ and $\sent$; $\query$ is set to be \post $\ut_{\numTurns}$.
%These methods utilize either a pre-trained BERT language model, denoted by \BERTpre, or a similar BERT model that was fine-tuned for sentence retrieval, denoted by \BERTfine  (see Section \ref{sec:BERT_finetuning}). Whenever we use a BERT model, either \BERTpre or \BERTfine, we take the pooler output vector of the BERT model as the single output, denoted by $\BERTout$.

\myparagraph{\origBERT} 
Inspired by a study of a few BERT \cite{BERT} architectures for ad hoc document retrieval \cite{Qiao+al:19a}, the Independent Representation method, \origBERT in short, uses a pre-trained BERT to produce two vectors: (i) for the query ($\ut_{\numTurns}$), and (ii) for the sentence $\sent$. Specifically, the inputs to BERT for generating the two vectors are ``[CLS] $\ut_{\numTurns}$ [SEP]'' and ``[CLS] $\sent$ [SEP]'', respectively. The output vectors, which correspond to contextual top-level representation of the [CLS] token, and which are denoted $v^{\ut_{\numTurns}}$ and $v^{\sent}$, respectively, are used to compute the retrieval score of $\sent$ via cosine similarity, $Cosine(v^{\ut_{\numTurns}},v^{\sent})$. 

\myparagraph{\rankBERT{X}} 
%This method takes an encoder-only Transformer, BERT \cite{Delvin+al:18a}, with input ``[CLS] $\query$ [SEP] $\sent$ [SEP]''. Following \cite{nogueira2019passage}, we train BERT whose output is fed into a softmax function that predicts a relevance score for sentence $\sent$ to fit query $\query$:
Following Nogueira et al. \cite{nogueira2019passage}, the \rankBERT{X}
method is a BERT model which is fine-tuned on a retrieval
task using dataset X. \rankBERT{X} produces a relevance score for sentence $\sent$
with respect to the query $\ut_{\numTurns}$ through a softmax
layer. It gets ``[CLS] $\ut_{\numTurns}$ [SEP] $\sent$ [SEP]'' as an
input, and outputs a vector $\outvec$, which corresponds to the
contextual top-layer representation of [CLS]. The output $\outvec$ is
then fed into a softmax function to estimate the relevance score for $\sent$:
\begin{equation}
  \label{eq:relScore}
    \score{\sent;\ut_{\numTurns}} = Softmax(\arbmat_{score}\ \outvec + \arbvec_{score});
\end{equation}
$\arbmat_{score}$ and $\arbvec_{score}$ are learned parameters.

\input{distant_supervision}

%% file: distant_supervision.tex
\subsection{A Weakly Supervised Training Set}
\label{sec:distant_supervision}

As discussed in Section \ref{ssec:weak_sup_rel}, a common practice for
training neural retrieval models is to further fine-tune a ranking
model for the end task.
%, and then further fine tune it on a training set that corresponds to the test set to be tested\hagai{cite?}.
Since obtaining large scale manually annotated training set is expensive and time consuming, we
%follow prior works that
use weak supervision instead.
%(see Section \ref{ssec:weak_sup_rel}).
To this end, we propose an automated method for generating pseudo relevance labels for sentences with respect to %Reddit conversations.
conversations created from Reddit. The constructed training set is used in our experiments to fine-tune the \rankBERTArg{X} neural retrieval model in a weakly supervised manner.

As training data, we considered every thread that includes \\ Wikipedia grounded \posts, each considered as a \emph{\target}, within the training part of our Reddit collection (see Section \ref{ssec:collection}), resulting in 766,334 threads.
%As training data, we considered every \post in thread within the training part of our Reddit collection (see Section \ref{ssec:collection}). We only keep threads that include Wikipedia grounded \posts, each considered as a \emph{\target}, resulting in 766,334 examples. 
We filtered out threads with \targets of 5 words or shorter, which we found difficult to automatically label.
For each training example (thread), the initial sentence retrieval method (Section \ref{sec:initial_ranker}) is used to retrieve 1000 sentences. We only retained
%examples
threads for which at least one retrieved sentence appears in the Wikipedia sections linked by the Wikipedia grounded \target.  
Last, we observed that in threads in which the \target is followed by more \posts in the thread, which we refer to as \emph{future \posts}, the Wikipedia section linked in the \target was of more use within the \target itself.  We therefore kept only threads in which the \target is followed by additional \posts in the thread, ending with $72,953$ threads for training. We refer to these threads as \emph{Reddit conversations}.

Next, each candidate sentence is assigned a pseudo relevance label using a weak annotator.
To this end, we rely on a real example of natural continuation of the thread: the \target and the Wikipedia sections that the author indicated as helpful in writing it. If the \post's author referenced a specific Wikipedia section but not the entire document, we view this as an indication that other paragraphs are less relevant. We note that this assumption does not hold with high confidence for other documents not referenced by the \post's author, because the author might not be familiar with them.
Therefore, our weak annotator labels only sentences in the documents containing the pointed sections.

Unlike inference time, where only the conversation history is available, during training set construction the \target and the future \posts are also accessible and can be used to help the pseudo relevance labeling process. Let $\dialog$ of length $m$ denote a conversation composed of three parts: (a) the conversation history $\ut_{1}$ \ldots, $\ut_{\numTurns}$, (b) the \target $\ut_{\numTurns+1}$, and (c) the future turns $\ut_{n+2}$ \ldots, $\ut_{m}$.
To induce pseudo relevance labels using these \posts, we used Reciprocal Rank Fusion (RRF) \cite{cormack2009reciprocal} to fuse the scores assigned by four retrieval
methods:
%The most robust
%%-OK: we cannot say it
%\hagai{Can we say it without showing corresponding results?} weak annotator was obtained 
%
%by using
%Reciprocal Rank Fusion (RRF) \cite{cormack2009reciprocal} was used.  ranking me%thods:

\myparagraph{Term-based cosine similarity}
Cosine similarity between the TF-IDF vectors of a sentence and the \target $\ut_{\numTurns+1}$. We use the RSJ IDF version \cite{Robertson+Walker} with stopword removal and Krovetz stemming.%\hagai{cite?}  

\myparagraph{BERT-based cosine similarity}
Similarly to \origBERT (see Section \ref{ssec:eval_models}), we use a pre-trained BERT \cite{BERT} to separately embed the candidate sentence $\sent$ to $v^{\sent}$ and the \target $\ut_{\numTurns+1}$ to $v^{\ut_{\numTurns+1}}$. We then compute the cosine similarity
$Cosine(v^{\ut_{\numTurns+1}},v^{\sent})$.
%between their respective output (top layer) embedding vectors of the CLS token.

%\myparagraph{Fused Initial Ranker}

\myparagraph{Fused LM}
%Let denote $\retList_{h}$ a mixture of language models induced from the \posts in the conversation history. The language models are increasingly weighted with the exponential decay function ${\alpha_i}$ from Section \ref{sec:docRank} where $T=n$. Let denote $\retList_{f}$ a mixture of language models induced from future \posts, where the ${\alpha_j}$'s (we use a different notations to indicate that these are different weights since the number of future \posts might be a different from the number of history \posts) are in reverse order, so the closer the \post to the \target the more relevant it is. $\retList_{t}$ is denoted as the language model of the \target. We score a sentence ~\sent~ using $-\ce{\mleProb{k}{\cdot}}{\dirLM{\sent}{\cdot}}$, resulting in a ranking score of ~\sent~ in ranked list $\rank{\retList_{k}}{s}$ induced by the \posts from $\retList_{k} \in  \set{\retList_{h},\retList_{t},\retList_{f}}$ 
Let  $\mixProb{\history}{\arbTerm} \definedas \sum_{i = 1}^\numTurns \alpha_i \mleProb{\ut_i}{\arbTerm}$
%denote a mixture of language models induced from the \posts in the conversation \emph{history}.
denote the probability assigned to $\arbTerm$ by a mixture of language models induced from the \posts in the conversation \emph{history}.
The language models are increasingly weighted with the exponential decay function ${\alpha_i}$ from Section \ref{sec:docRank} where $T=n$ and $\indSet = \{1, ..., \numTurns\}$.
Similarly, let $\mixProb{\future}{\arbTerm} \definedas \sum_{i = \numTurns+2}^m \alpha_i \mleProb{\ut_i}{\arbTerm}$ denote the probability assigned to $\arbTerm$ by a mixture of language models induced from \emph{future} \posts, where $T = \numTurns+2$ and $\indSet = \{\numTurns+2, ..., m\}$ for the ${\alpha_i}$'s. %are in reverse order, so the closer the \post to the \target the more relevant it is. %\mleProb{\ut_{\numTurns+1}}{w} is denoted as the language model of the \target.
We score a sentence $\sent$ using $-\ce{q(\cdot)}{\dirLM{\sent}{\cdot}}$, where $q(\cdot) \in \{\mixProb{\history}{\cdot}, \mixProb{\future}{\cdot}, \mleProb{\ut_{\numTurns+1}}{\cdot}\}$, resulting in the ranked lists $\retList_\history, \retList_\future$ and $\retList_{\ut_{\numTurns+1}}$.
Let $\rank{\retList_j}{\sent}$ be the rank of sentence $\sent$ in $\retList_j$; the highest rank is $1$.
The final score of $\sent$ is:
%is computed with weighted RRF:  
\begin{equation}
\label{eqn:weaklyFusedList}
Score(s) = \frac{\lambda}{2}\frac{1}{\nu +\rank{\retList_{h}}{s}} + (1-\lambda)\frac{1}{\nu +\rank{\retList_{t_{n+1}}}{s}} + \frac{\lambda}{2}\frac{1}{\nu +\rank{\retList_{f}}{s}},
\end{equation}
where $\lambda$ and $\nu$ are free parameters.

\myparagraph{Fused BERT} 
We utilize a fine-tuned \rankBERTArg{X} from Section
\ref{ssec:eval_models} to rank the sentences against each of the $m$ \posts in conversation $\dialog$. As a result, we obtain $m$ ranked lists $\retList_i, i \in \{1, ..., m\}$; $\rank{\retList_i}{\sent}$ is the rank of sentence $\sent$ in $\retList_i$.
Let 
%$rank_{\retList_{h}}$ 
$\retList_{\history}$ denote
a list fused from the ranked lists induced by the \posts in the conversation history. The fusion score of $\sent$ is $\sum_{i=1}^\numTurns \alpha_i \frac{1}{\nu + \rank{\retList_{i}}{\sent}}$, where the $\alpha_i$'s are computed with $T=n$ and $\indSet = \{1, ..., \numTurns\}$.
%The fusion was applied using an increasingly weighted RRF with the exponential decay function ${\alpha_i}$ from Section \ref{sec:docRank} where $T=n$.
Similarly, we  
%$rank_{\retList_{f}}$
create a list $\retList_{\future}$ by fusing the ranked lists induced by the future \posts ($i \in \{\numTurns+2, ..., m\}$) where $T = n+2$ and $\indSet = \{\numTurns+2, ..., m\}$ for the ${\alpha_i}$'s.
%are in reverse order.
%$rank_{\retList_{t}}$
%$\retList_{\ut_{\numTurns+1}}$
%is denoted as the ranked list induced from the \target.
Finally, we assign each sentence $\sent$ a score which results from fusing
%$rank_{\retList_{h}}$, $rank_{\retList_{t}}$ and $rank_{\retList_{f}}$
$\retList_{\history}$, $\retList_{\ut_{\numTurns+1}}$ and $\retList_{\future}$ as in Eq. \ref{eqn:weaklyFusedList}.

%We utilize \rankBertArg{X} from Section
%\ref{ssec:eval_models} to rank the sentences against each \post in the
%dialogue, including the history \posts, the \target and future \posts;
%thus, we obtain $m$ ranked lists: $\retList_{t_1},...,\retList_{t_m}$. We
%then fuse the lists for the history \posts and future \posts
%separately using the decay factor weights $\alpha_i$ from Section
%\ref{sec:initial_ranker}, resulting in the three ranked lists
%$\retList_{h}$, $\retList_{t}$ and $\retList_{f}$, which are then
%fused as in Eq. \ref{eqn:weaklyFusedList}.

 \

\begin{figure}[t]
    \makebox[1 \columnwidth][c]{
    \includegraphics[width=1\columnwidth, height=7cm]{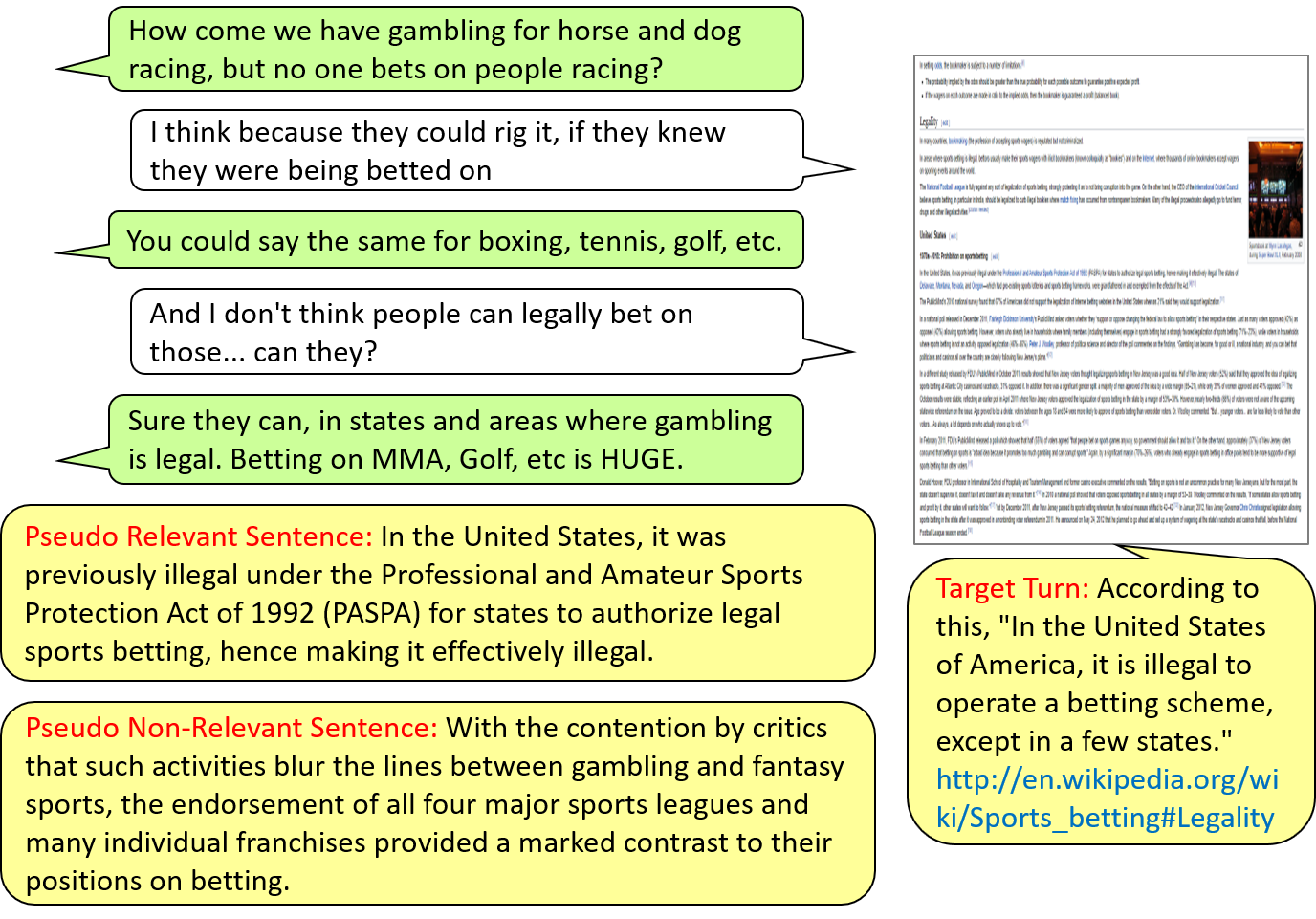}
    }
    \caption{Training example of a conversation created from Reddit with pseudo relevant and non-relevant sentences.}
    \label{fig:conv-example-train}
\end{figure}

Once all sentences in a document with a referred section are assigned
a retrieval score, the $k$ sentences in the {\em pointed section} with
the highest retrieval scores are labeled as pseudo relevant; the $k$
sentences in the {\em document} with the lowest retrieval scores,
excluding sentences in the pointed section, are labeled as pseudo
non-relevant. This selection of pseudo non-relevant sentences strikes a balance
between the fact that the sentences might be topically related to the conversation by the virtue of being part of the same document, and the lower likelihood of their relevance due to being the bottom ranked in the document. Figure \ref{fig:conv-example-train} exemplifies an outcome
of the auto-labeling algorithm.

%% file: experiments.tex
\section{Experiments}
\label{sec:experiments}

We evaluated the models presented in Section \ref{sec:models} on our novel testset described in Section \ref{ssec:test set}. We next describe the setting and the results of these experiments.

\input{exp_setting}

\input{results}

%% file: exp_setting.tex
\subsection{Experimental Setting}
\label{sec:expSetting}

\subsubsection{Data Split and Metrics}
\label{ssec:metrics}
 \
 
We randomly split the the test set dialogues (Section \ref{ssec:test set}) $50$ times into two equal-sized sets: validation set, for hyperparameter tuning, and test set. In each split, the number of grounded and ungrounded dialogues is equal between the validation and test sets. 

We measure Mean Average Precision (\map), NDCG of the $5$ highest ranked sentences (\ndcgfive) and Mean Reciprocal Rank (\mrr)
of the highest ranked relevant sentence. 
The results we report are average and standard deviation over the $50$ test sets.
Statistically significant performance differences are determined using the two tailed permutation (randomization) test with $10,000$ random permutations and $p \le 0.05$. 
The tests are performed with respect to the performance attained for each of the $50$ test sets. We use Bonferroni correction for multiple hypothesize testing.

\subsubsection{Initial Ranker}
\label{ssec:initial_ranker}
 \
 
As document collection we use the Wikipedia dump from 2020-01-01\footnote{\url{https://dumps.wikimedia.org/enwiki/20200101/enwiki-20200101-pages-articles-multistream.xml.bz2}}. Parsing was mainly done by Wikiextractor\footnote{\url{https://github.com/attardi/wikiextractor.git}} with a few manual improvements. We indexed with Indri\footnote{\url{http://www.lemurproject.org/indri}}.

We use the \initRank (Section \ref{sec:initial_ranker}) with Krovetz stemming for documents and dialogues, and stopword\footnote{\url{http://www.lemurproject.org/stopwords/stoplist.dft}} removal only for dialogues. We did not include candidates that become empty after stopword removal. We retrieve the top-1000 documents for each dialogue and the top-50 sentences for each retrieved document.
%All hyperparameters were fixed empirically \hagai{Another word?} on a small subset of examples %\hagai{in what value ranges? on which metric?}; 
We set $\beta$ to 0.3 in Eq. \ref{eq:dcoRepDiag} and \ref{eq:sentRepDiag}, $\gamma$ to 0.75 in Eq. \ref{eq:final_score}, and the Dirichlet prior was set to 1000 \cite{Zhai+Lafferty:01a}.
%In Eq. \ref{eqn:timeBasedWeights},
For computing $\alpha_i$ (Section \ref{sec:docRank}),
${\delta}$ was set to 0.01, following \cite{Bruce+Xiaoyan-Time-Based}.

\subsubsection{Training Settings} 
\label{sec:BERT_finetuning}

In all experiments, BERT-Large \cite{BERT} was used as the pre-trained BERT model.
We experimented with two trained \rankBERT{X} variants with the same architecture (Section \ref{ssec:eval_models}). The first, denoted \rankBERT{\BERTfineMS}, takes a pre-trained BERT model and fine tunes it on the passage retrieval task in MS Marco \cite{nguyen2016ms}. The input is ``[CLS] $q$ [SEP] $p$ [SEP]'', where $q$ is the query and $p$ is the passage in the MS Marco dataset. The model is trained with a pointwise classification loss\footnote{Training with pairwise loss showed no improvement.} \cite{nguyen2016ms}. The second variant,\\ \rankBERT{\BERTfineR}, is a version of \rankBERT{\BERTfineMS} further fine-tuned, end-to-end, using our weakly supervised training set (Section \ref{sec:distant_supervision}), with input as described in Section \ref{ssec:eval_models}. 
We note that this two stage fine-tuning of BERT-based rankers was also explored in \cite{li2020parade}.

% \rankBERT was evaluated with either \BERTfineMS or \BERTfineR, including the trained ranking layer, without further training. Specifically, we trained ~\rankBERT~ end-to-end on our weakly supervised training set, with \BERTfineMS %\hagai{Didn't we start from \BERTfineMS?} 
% as a starting point. This practice allows us to evaluate the merits of our weak supervision training set.

\subsubsection{Hyper-Parameter Tuning.}
\label{ssec:hyperparams_tuning}
 \
 
All hyper-parameters were tuned on the validation sets with \map as the optimization criterion.
%We provide a list of parameters and the value range tested for each.
We list the hyper-parameters and their corresponding tested value ranges.
The Dirichlet smoothing parameter used in LM, %\crossEnt, %\hagai{and also in the CE baseline?}) 
$\mu$, is set to values in $\set{1000,2000}$~\cite{Zhai+Lafferty:01a}. 
For Okapi BM25, we used $k_1 \in \set{1.2, 2, 4, 8, 12}$ and $b \in \set{0.25, 0.5, 0.75, 1}$.

We fine-tuned a pre-trained BERT on MS Marco using the Adam optimizer with learning rate $\in \set{3e-6,3e-8}$ and 
batch size $\in \set{64,128}$. As in \cite{nogueira2019passage}, we trained on $12.8$M query-passage pairs from the dataset.
We further fine-tuned \rankBERT{\BERTfineMS} with the same training scheme for $10$ epochs on our weakly supervised training set, with the hyper-parameter values mentioned above, to obtain \rankBERT{\BERTfineR}.
%All \rankBERT variants were trained for 10 epochs on Google's TPU\footnote{\url{https://cloud.google.com/tpu/}} v3-8 and the best model snapshot was chosen based on the validation set. We trained ~\rankBERT~ using the Adam optimizer with learning rate $\in \set{3e-6,3e-8}$ and batch size $\in \set{64,128}$. 
%\idan{I removed a sentence about training \cite{nogueira2019passage} model, please return if needed but better explain.}
The maximum sequence length is 512. All \posts were truncated to 70 tokens, which is the maximum length of \posts in the test set, affecting less than $0.1\%$ of the training set candidates.

For automatic labeling of the training set (Section \ref{sec:distant_supervision}), 
%we did not tune hyperparameters. 
we used RRF with default parameter $\nu$=60, empirically set $m$ future \posts to 4, $\lambda$ to 0.3 in Eq. \ref{eqn:weaklyFusedList}, and %decay rating to $0.01$ 
and $\delta$ to $0.01$, when computing ${\alpha_i}$
%in Section \ref{sec:docRank},
for the \emph{Fused LM} 
%\emph{Fused Initial Ranker} 
and \emph{Fused BERT} methods.
We select the top-$3$ and bottom-$3$ sentences to serve as the pseudo relevant and non-relevant sentences, respectively.
%Parameter $k$ was set to 3 when we selected the top $k$ pseudo relevant and bottom $k$ pseudo non-relevant sentences.
In BERT-based weak annotators, \posts and sentences were truncated to contain $64$ and $112$ tokens, respectively. %\hagai{This passage presents hyper parameters that weren't really tuned right?}

%% file: results.tex
\begin{table}[t]
\small
  \caption{\label{tab:main-results} Model performance on the testset. '$i$' and '$r$' mark statistically significant differences with the \initRank and \rankBERT{\BERTfineR}, respectively. Boldface marks the best performance in a column.}  

    \input{tables/main-results}

\end{table}

\begin{table}[t]
\small
  \caption{\label{tab:diff-results} MAP performance for ungrounded and Wikipedia grounded dialgogues. '$i$' and $'r$ mark statistically significant differences with the \initRank and \rankBERT{\BERTfineR}, respectively. Boldface marks the best result in a column.}

    \input{tables/diff-results}

\end{table}

% \iffalse
% \begin{table}[t]
% \small
%   \caption{\label{tab:val-distant} The effectiveness of using our weak supervision approach. Models in the first and second row in a block use BERT which was fine-tuned on MS MARCO (\BERTfineMS) and further fine tuned using our weak supervision approach on Reddit (\BERTfineR), respectively. Statistically significant with the first row in each block is marked with m.}
% \label{tab:val-distant}
% %\input{tables/val_distant_old_order.tex}
% \begin{tabular}{|l|c|c|c|}
% \hline
%  & \map & \ndcg  & \mrr \\
% \specialrule{.2em}{.1em}{.1em}
% \rankBERT{}(MS) & $.328^{\pm .008}$ & $.457^{\pm .012}$ & $.444^{\pm .012}$ \\
% \hline
% \rankBERT{}(MS $\rightarrow$ R) & $\mathbf{.345}^{\pm .009}_{m}$ & $\mathbf{.480}^{\pm .013}_{m}$  & $\mathbf{.461}^{\pm .012}_{m}$  \\
% \hline
% \end{tabular}

% \end{table}
% \fi

\subsection{Results}
\label{sec:results}
%\hagai{Worth to show the performance difference between types of dialogues}\itay{Doesn't Table 1 show that? I'm not sure if the MAP performance is also needed here.}
%\hagai{No, I mean the metrics MAP, etc.}\itay{Will do. I'll use 200 for validation and 200 for test (instead of 400 for each as done for Table 2).} %With respect to which model? We didn't tune the hyperparameters only for the initial ranker and I'm not sure if the initial ranker is interesting for showing the differences; note that we can't present the MAP for the others as we don't have validation sets to tune their hyperparameters.}

The performance of all evaluated models on the proposed dataset is presented in Table \ref{tab:main-results}. We see that %all models that considered only the last turn $\ut_{\numTurns}$ as the query to be matched with the candidate sentence  -- 
\origBERT, LM %\crossEnt 
and \okapi, which only match the last turn $\ut_{\numTurns}$ with a sentence perform statistically significantly worse than
%which considered only the last turn $\ut_{\numTurns}$ as the query to be matched with the candidate sentence, performed significantly worse than
%models
the \initRank
%that also matched the sentence with the conversation history.
that matched the sentence with the entire dialogue history.
%These models, which rerank the the \initRank's output, deteriorated the \initRank's ranked output.
This is a clear evidence that using the dialogue history is important to effectively represent the information need behind the last turn. This finding echoes a similar insight from prior work on response selection for dialogues \cite{Yan+al:16a}. It is surprising to see that pre-trained BERT (\origBERT) underperforms compared to the token-based language models: LM %\crossEnt 
and \okapi. This indicates the importance of fine-tuning BERT models to ranking tasks.
%models of \crossEnt and \okapi, indicating the necessity of fine-tuning BERT models to ranking tasks.

However, once fine-tuned for a retrieval task, a BERT model statistically significantly outperforms all token-based retrieval methods: compare the performance of \rankBERT{X} methods to that of the other methods in Table \ref{tab:main-results}.
% as seen by the significantly higher performance of \rankBERT{X} variants compared to the \initRank.
This shows that once the dialogue context is taken into consideration in the initial sentence retrieval, re-ranking can improve results with fine-tuned models even when only the last turn and the sentence are matched. In future work, we plan to investigate whether neural models that consider also the dialogue history can further improve performance as indicated in some prior work on conversational search \cite{Gao+al:22a,Zamani+al:22a}.
%conversation history and the document context of the retrieved sentence can further improve these results.

Table \ref{tab:main-results} also shows the performance gain of further fine tuning a model on the specific task at hand. Indeed, while \rankBERT{\BERTfineMS} outperforms all non-fine-tuned models, the \rankBERT{\BERTfineR} model, which was further fine-tuned using our weakly supervised training set,
improves the performance. This method attains the highest performance with all performance gains over other methods being statistically significant.
%improves the results, achieving the (statistically significant) highest performance in this experiment.
This finding also demonstrates the effectiveness of our weak annotator and weakly supervised training set, showing that performance can be improved without manual annotation for training.

To offer more insight on the types of dialogues in our testset, we computed the MAP of the tested models only on Wikipedia grounded dialogues and only on ungrounded dialogues (see Section \ref{ssec:test set}). The performance results, presented in Table \ref{tab:diff-results}, show that all models perform better on the Wikipedia grounded dialogues; yet, the relative performance order of methods for ungrounded and grounded diagloues is the same\footnote{Results for MRR and NDCG@5 show the same patterns as for MAP, and are omitted as they convey no additional insight}.
%The results are shown for the MAP measure, but this is also true for the \mrr and \ndcg measures.
%This indicates that Reddit conversations that include references to Wikipedia have structures that retrieval models find somewhat easier to find relevant sentence for. In future work we plan to investigate whether different model architectures should be applied to each conversation type.
Thus, we conclude that Reddit conversations that include references to Wikipedia have structure, and embody information, that allow for more effective retrieval than conversations with no such references.
%This indicates that Reddit conversations that include references to Wikipedia have structures that retrieval models find helpful when distinguishing between relevant and non relevant sentences, resulting in relevant sentences ranked higher.
In future work we plan to investigate whether different model architectures should be applied to each conversation type.

% that \rankBERT{} is significantly better than all other models, both the one that was fine tuned on MS-MARCO and the one that was further fine tuned on Reddit dialogues with our weekly supervised training method (Sec. \ref{sec:distant_supervision}).
% When comparing \rankBERT{}(\BERTfineMS) and \rankBERT{}(\BERTfineR), it is also shown that training with our pseudo labels significantly improve performance,  verifying the strength of our weakly supervised training method %\hagai{Can we measure also significance for \BERTfineMS? (if we leave it in this table)}.

% Another observation is the power of the initial ranker, which leverage the entire dialogue context, compared to other lexical models (i.e., BM25 and CE) as well to \origBERT, which rely only on the last \post. But still Contextual LMs that were fine tuned for passage re-ranking are better.

%% file: tables/main-results.tex
\begin{tabular}{|l|c|c|c|}
\hline
 & \map & \ndcg  & \mrr \\
\specialrule{.2em}{.1em}{.1em}
\initRank & $.238^{\pm .007}$ & $.355^{\pm .012}$  & $.353^{\pm .012}$ \\
\hline
LM & $.185^{\pm .006}_{ir}$ & $.253^{\pm .012}_{ir}$ & $.256^{\pm .011}_{ir}$ \\ 
\hline
\okapi & $.185^{\pm .006}_{ir}$ & $.259^{\pm .010}_{ir}$ & $.258^{\pm .009}_{ir}$ \\ 
\hline
\origBERT & $.172^{\pm .004}_{ir}$ & $.236^{\pm .009}_{ir}$ & $.240^{\pm .008}_{ir}$ \\
\hline
\rankBERT{\BERTfineMS} & $.328^{\pm .008}_{ir}$ & $.457^{\pm .012}_{ir}$ & $.444^{\pm .012}_{ir}$ \\
\hline
\rankBERT{\BERTfineR} & $\mathbf{.345}^{\pm .009}$ & $\mathbf{.480}^{\pm .013}$  & $\mathbf{.461}^{\pm .012}$ \\
\hline
\end{tabular}

%% file: tables/diff-results.tex
\begin{tabular}{|l|c|c|c|}
\hline
& Ungrounded & Wikipedia Grounded \\
%\hline
\specialrule{.2em}{.1em}{.1em}
\initRank & $.223^{\pm .009}_{r}$ & $.252^{\pm .010}_{r}$ \\
\hline
LM & $.170^{\pm .009}_{ir}$ & $.197^{\pm .009}_{ir}$ \\ 
\hline
\okapi & $.168^{\pm .009}_{ir}$ & $.197^{\pm .009}_{ir}$ \\ 
\hline
\origBERT & $.159^{\pm .008}_{ir}$ & $.184^{\pm .008}_{ir}$ \\
\hline
\rankBERT{\BERTfineMS} & $.311^{\pm .012}_{ir}$ & $.340^{\pm .012}_{ir}$ \\
\hline
\rankBERT{\BERTfineR} & $\mathbf{.323}^{\pm .014}_{i}$  & $\mathbf{.362}^{\pm .012}_{i}$ \\
\hline
\end{tabular}

%% file: conc.tex
\section{Conclusions and Future Work}
\label{sec:conc}
We introduced the task of
sentence retrieval from a document corpus for open-ended dialogues. The goal is to retrieve sentences that contain information useful for generating the next \post in a given dialogue. Sentences that meet this criterion are deemed relevant to the dialogue.
%We deemed these sentences as relevant.

To evaluate retrieval models for the dialogue-based sentence retrieval task, we created a dataset consisting of 846 Reddit
dialogues and candidate retrieved sentences from Wikipedia. The dataset also includes human relevance judgments for each sentence. The dataset is available at: \url{https://github.com/SIGIR-2022/A-Dataset-for-Sentence-Retrieval-for-Open-Ended-Dialogues.git}.

We are not aware of other publicly available datasets suitable for the evaluation of {\em retrieval effectiveness} for open-ended dialogues. A unique characteristics of the task is the fact that there is no explicit statement of an information need in the form of questions or queries. Accordingly, the relevance definition we use for sentences is not based on satisfaction of an information need or on being an answer to a question. Rather, as noted above, it is based on the usefulness of the information included in the sentence for generating the next response in the dialogue.

We evaluated several retrieval models on the the novel dataset,
including (unigram) language modeling, probabilistic retrieval (Okapi BM25) and
neural rankers. To fine-tune neural rankers to the proposed open-dialogue retrieval task, we
presented a weak annotator that automatically assigns pseudo-relevance
labels to training set sentences.
We show that a
%neural ranker fined-tuned
neural ranker which was fined-tuned
using our weakly supervised training set outperforms
all other tested models, including a neural ranker
fine-tuned on the MS Marco passage retrieval dataset.

% Our evaluation shows the merits of utilizing the dialogue history, especially among the lexical models; the initial ranker significantly outperforms other strong lexical models. In future work, we would like to construct models that utilize the dialogue history and to evaluate their merits.

In future work, we would like to devise BERT-based retrieval models that are
trained based on weak supervision alone, using a pre-trained BERT,
%is sufficient to train a neural ranker from a
%pre-trained model like BERT,
without the need for large
%manually
annotated training sets like MS Marco. We would also like to ground
generative language models with our retrieval models and study the
conversations that emerge from such grounding.

%% file: acknowledgements.tex
\section*{Acknowledgements}

We thank the reviewers for their comments. This work was supported in part by a grant from Google.

%% file: appendix.tex
\section{Annotation Instructions}
\begin{figure*}[t]
\makebox[1 \textwidth][c]{
\includegraphics[width=12cm,height=20cm]{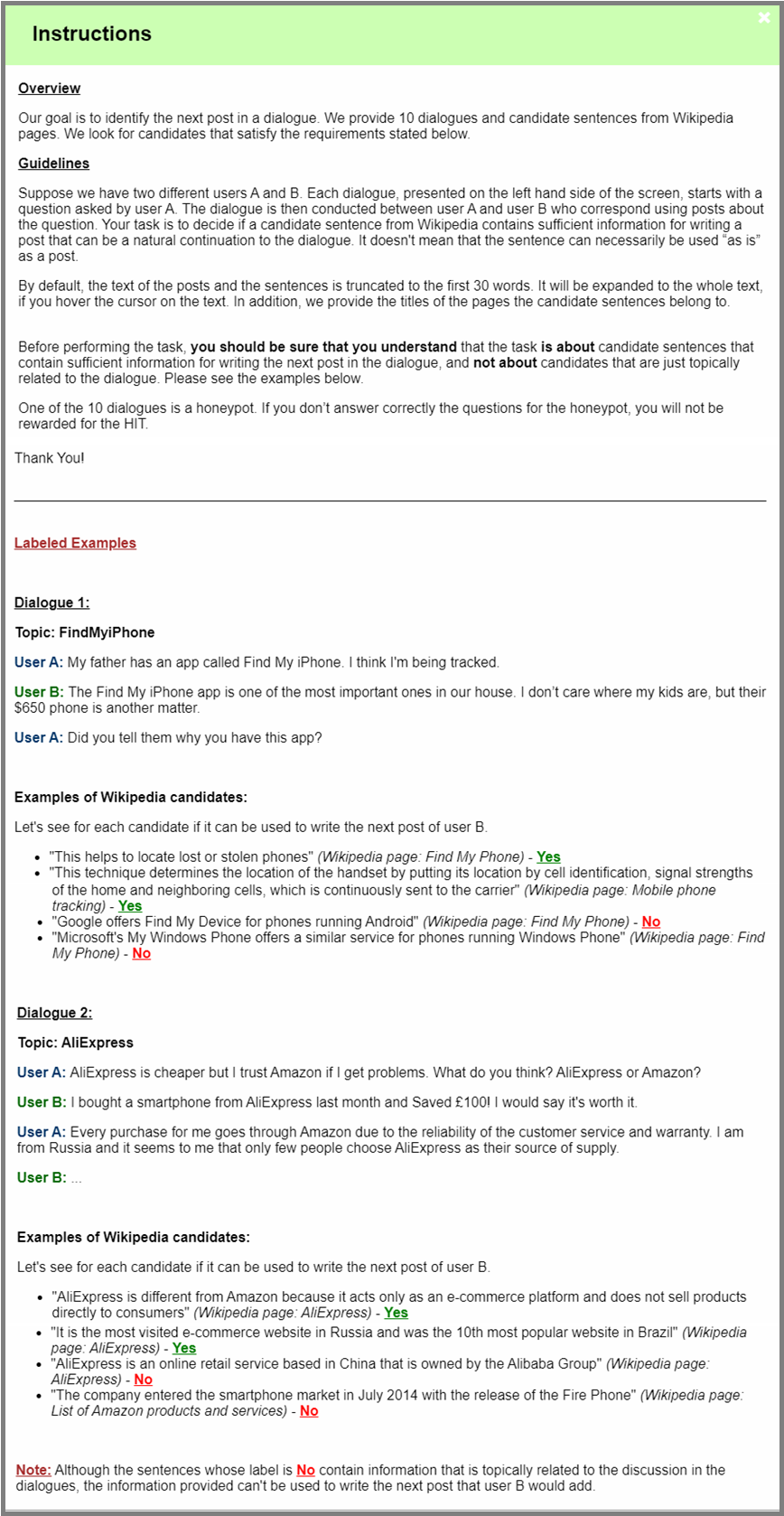}
}
\caption{\label{fig:mturk_guidelines} Test set annotation guidelines used in Mechanical Turk.}
\end{figure*}

Figure \ref{fig:mturk_guidelines} presents the instruction form provided to the annotators.

